\documentclass[aps,prl,reprint,groupedaddress]{revtex4-1}
\usepackage{amsmath,amssymb,graphicx}
\usepackage{color}

\begin{document}

\title{Observation of geometric parametric instability induced by the periodic spatial self-imaging of multimode waves}


\author{Katarzyna Krupa}\email[]{Corresponding author: krupa.katarzyna@yahoo.com}
\affiliation{Universit\'e de Limoges, XLIM, UMR CNRS 7252, 123 Av. A. Thomas, 87060 Limoges, France}
\author{Alessandro Tonello}
\affiliation{Universit\'e de Limoges, XLIM, UMR CNRS 7252, 123 Av. A. Thomas, 87060 Limoges, France}
\author{Alain Barth\'el\'emy}
\affiliation{Universit\'e de Limoges, XLIM, UMR CNRS 7252, 123 Av. A. Thomas, 87060 Limoges, France}
\author{Vincent Couderc}  
\affiliation{Universit\'e de Limoges, XLIM, UMR CNRS 7252, 123 Av. A. Thomas, 87060 Limoges, France}
\author{Badr Mohamed Shalaby}
\affiliation{Physics Department, Faculty of Science, Tanta University, Egypt}
\affiliation{Universit\'e de Limoges, XLIM, UMR CNRS 7252, 123 Av. A. Thomas, 87060 Limoges, France}
\author{Abdelkrim Bendahmane}
\affiliation{Universit\'e de Bourgogne Franche-Comt\'e, ICB, UMR CNRS 6303, 9 Av. A. Savary, 21078 Dijon, France }
\author{Guy Millot}
\affiliation{Universit\'e de Bourgogne Franche-Comt\'e, ICB, UMR CNRS 6303, 9 Av. A. Savary, 21078 Dijon, France }
\author{Stefan Wabnitz}
\affiliation{Dipartimento di Ingegneria dell'Informazione, Universit\`a di Brescia, and INO-CNR, via Branze 38, 25123, Brescia, Italy}


\begin{abstract}
Spatio-temporal mode coupling in highly multimode physical systems permits new routes for exploring complex instabilities and forming coherent wave structures.  
We present here the first experimental demonstration of multiple geometric parametric instability sidebands, generated in the frequency domain through resonant space-time coupling, owing to the natural periodic spatial self-imaging of a multimode quasi-continuous-wave beam in a standard graded-index multimode fiber. 
The input beam was launched in the fiber by means of an amplified microchip laser emitting sub-nanosecond pulses at 1064 nm.
The experimentally observed frequency spacing among sidebands agrees well with analytical predictions and numerical simulations. 
The first order peaks are located at the considerably large detuning of 123.5 THz from the pump. These results open the remarkable possibility to convert a near-infrared laser directly into a broad spectral range spanning visible and infrared wavelengths, 
by means of a single resonant parametric nonlinear effect occurring in the normal dispersion regime.
As a further evidence of our strong space-time coupling regime, we observed the striking effect that all of the different sideband peaks were carried by a well-defined and stable bell-shaped spatial profile.

\end{abstract}


\maketitle


Pattern formation as the result of parametric instability (PI) is an universal phenomenon that is widely encountered in many branches of physics \cite{HohenbergRMP}. PIs emerge in wave propagation thanks to the interplay between nonlinearity and the dispersion of the medium, when one of the medium parameters is periodically modulated along the longitudinal direction. In the case of externally forced systems, PI is commonly referred to as the Faraday instability, following its initial observation in hydrodynamics under the external modulation of the vertical position of an open fluid tank \cite{Faraday1831}. Besides fluid mechanics, Faraday-like patterns were subsequently reported in a variety of physical contexts such as crystallization dynamics, chemical systems or laser physics \cite{ErneuxFarad,Staliunas2013PRA,Saarloos1995,Lin2000,Perego2015}. In addition to parametrically forced systems, many physical systems naturally exhibit collective oscillations that may lead to a so-called geometric-type of parametric instability (GPI). Both Faraday-like and geometric instabilities can be found, for instance, in Bose-Einstein condensates, where they can be induced either by the harmonic modulation of the nonlinear interaction \cite{EngelsPRL,JibbouriBEC,Staliunas2002}, or by the profile of the trapping potential \cite{JibbouriBEC}, respectively.
  
In nonlinear optics, the propagation of an intense continuous wave (CW) may undergo modulation instability (MI), which leads to the exponential amplification of spectral sidebands at the expense of the CW. In scalar optical fiber propagation, MI is observed in the anomalous dispersion regime only \cite{Agraval1995NonlinearOptics}.  MI may also be observed in the normal dispersion regime by exploiting, e.g., higher-order dispersion \cite{Millot2003} or vector propagation in birefringent fibers \cite{wab88a}.  On the other hand, PIs are observed in the presence of longitudinal periodic forcing, such as periodic amplification \cite{matera}, or dispersion \cite{Smith1996R34, Abdullaev1996R37, Consolandi2002R36,Droques12,Finot}. 

Nonlinear optics also offers examples of GPIs. 
The periodic energy exchange between a fundamental wave and its second harmonic in crystals with quadratic nonlinearity may induce a spatial parametric instability of geometric nature \cite{TrilloWabSteg}.
More recently, and of relevance beyond nonlinear optics, it was shown that the natural periodic evolution typical of Fermi-Pasta-Ulam recurrence is also at the origin of GPI \cite{WabWet}.

It has been recently demonstrated that multimode wave coupling in nonlinear optical fibers generates complex spatiotemporal dynamics that cannot be fully represented within the usual approach of separating the time domain from the transverse spatial coordinates. Although multimode fibers (MMFs) have been commercially available since the 1970s, it is only very recently that nonlinear mode coupling in MMFs has attracted a renewed research interest. From the perspective of fundamental physics, MMFs provide a perfect test-bed for advancing our understanding of complex spatiotemporal dynamics, and for studying many intriguing nonlinear spectral shaping effects \cite{Wright2015R31, Picozzi2015R30}. A very recent paper by Wright et al. \cite{Wright2015R29} reported the generation of a discrete set of dispersive waves at visible wavelengths, induced by the mechanism of quasi-phase-matching (QPM) with the respective harmonics of spatiotemporal multimode soliton (MMS) oscillations. The MMS was produced by injecting a femtosecond pump pulse in the anomalous dispersion regime of a graded-index multimode fiber (GRIN MMFs). 

In a GRIN MMF, the equal spacing of the modal wave numbers provides a natural periodic self-imaging effect for a multimode CW beam that propagates in the otherwise longitudinally uniform medium. In this work, we experimentally demonstrate for the first time that the natural periodicity of CW beams enables the observation of GPI sidebands in the normal dispersion regime, where MI leading to temporal break-up of the CW into a set of MMS cannot take place.

The present type of parametric instability is fundamentally different from all other known types of instabilities. First theoretically predicted by S. Longhi \cite{Longhi2003R27}, GPI in a MMF requires space-time coupling. When exciting a large number of modes in a standard parabolic GRIN MMF, the multimode beam propagates through the fiber and it experiences longitudinal spatial oscillations with the period $\xi=\pi\rho/\sqrt{2\Delta}$, where 
$\Delta=(n_{co}^2-n_{cl}^2)/2n_{co}^2$ is the relative index difference, $\rho$ is the fiber core radius, $n_{co}$ is the maximum core refractive index and $n_{cl}$ is the cladding refractive index. 
The wave-vectors of pump $k_P$, and of the GPI Stokes $k_S$ and anti-Stokes $k_A$
sidebands are expected to satisfy the QPM condition $2k_P-k_S-k_A=-2\pi h/\xi$, where $h=1,2,3,...$. If we limit the frequency dependence of the refractive index to the presence of group velocity dispersion at the pump wavelength ($\kappa''$), the generated GPI sidebands
are detuned from the pump by the discrete set of resonant frequency offsets $f_h$ satisfying the condition
$(2\pi f_h)^2=2\pi h/(\xi\kappa'')-2n_2\hat{I}\omega_0/(c\kappa'')$, 
where $n_2$ is the nonlinear Kerr index, $\omega_0$ is the angular frequency of the pump,   
and $\hat{I}$ is the path-averaged intensity of the beam. Following Ref.\cite{Longhi2003R27}, the condition $h=0$ reduces to the well-known MI of a uniform or non-oscillating CW beam, which requires anomalous group velocity dispersion $\kappa''<0$. 

In the normal dispersion regime, GPI occurs for $h=1,2,3,\dots$.
In our experiment, resonant frequencies depend only weakly upon the quasi-CW beam intensity: this permits to approximate well their values as $f_h\simeq\pm\sqrt{h}f_m$, where $2\pi f_m=\sqrt{2\pi/(\xi\kappa'')}$. 
With a standard GRIN MMF with a $\rho=26$ $\mu m$, $n_{co}=1.470$, $n_{cl}=1.457$ and $\kappa^ {''}=16.55\times 10^{-27}s^2/m$ at the pump wavelength of 1064 nm, $\Delta=8.8\times 10^{-3}$, so that the self-imaging period is $\xi=0.615$ $mm$. These values lead us to analytically predict an extremely large frequency detuning for the first resonant sideband, namely $f_1=f_m\simeq 125THz$. Such sideband shift is much larger than the value (8 THz) that was predicted by Longhi \cite{Longhi2003R27}, owing to the different fiber parameters and pump wavelength. However, as we will see, our theoretical prediction agrees well with numerical simulations (124.5 THz) and, more importantly, with the experimental value of $f_m\simeq123.5$ THz. 

Such large frequency shifts, linking directly the near infrared with the visible spectral domain, can be obtained in the normal dispersion regime by means of other parametric processes. For example, consider second harmonic generation in crystals \cite{Bortz1994R16, Zhou2014R37}, or intermodal four wave mixing (FWM) in optical fibers \cite{Poletti08,Tonello2006R33}, including the case of GRIN fibers \cite{Hill1981}. Although many FWM effects can be phase matched in a multimode propagation, none of them have shown the direct generation of a sequence of bright sidebands combined with such a large frequency shifts from the pump.

In the experiment, we used a 6m-long standard GRIN MMF with 52.1$\mu$m core diameter and 0.205 NA.
We pumped the fiber by an amplified Nd:YAG microchip laser, delivering 900 ps pulses at 1064 nm with the repetition rate of 30 kHz.
The polarized Gaussian pump beam was focused at the input face of the fiber with a FWHMI diameter of 35 $\mu$m, which was close to the value of the fiber core diameter. This implies that in the presence of unavoidable linear mode coupling, due to mechanical waveguide perturbations (bending radius of $\sim$ 120 mm) and technological fiber imperfections, we could excite a large number of guided modes (i.e., more than 200 including the polarization degree of freedom).

\begin{figure}
\includegraphics[width=\linewidth]{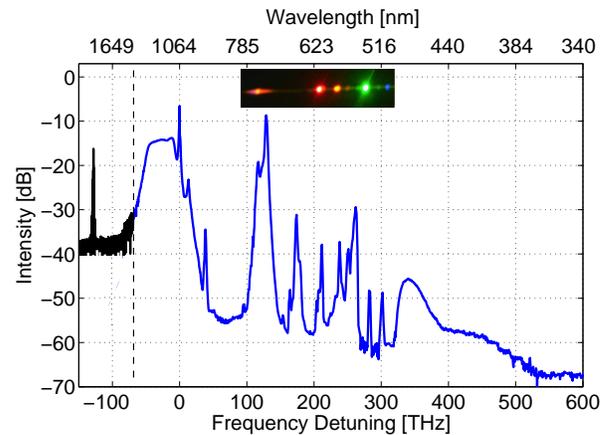}
\caption{Experimental spectra obtained in 6m-long GRIN MMF with 50 kW of input power $P_{p-p}$ by using two different spectrum analyzers; Inset: photographic image of the visible part of the spectrum dispersed by a diffraction grating.  \label{fig:1}}
\end{figure}

In order to numerically reproduce the GPI, self-induced by spatiotemporal nonlinear coupling, we solved the generalized $(3+1)$D nonlinear Schr\"odinger equation (NLSE) for the complex field envelope $A(x,y,t)$ [$\sqrt{W}/m$] (such equation is also referred to as the Gross-Pitaevskii 
equation) \cite{Longhi2003R27, Renninger2012R31, Yu1995R35, Mafi2012R36, Longhi04}:
\begin{equation}
\frac{\partial A}{\partial z}-i\frac{1}{2k_0}\nabla_{\bot }^2 A+i\frac{\kappa^{''}}{2}\frac{\partial^2 A}{\partial t^2}+i\frac{k_0\Delta}{\rho^2}r^2A=i\gamma|A|^2A,
\label{eq:1}
\end{equation}
where $k_0=\omega n_{co}/c$, $\gamma=\omega_0n_2/c$ is the fiber nonlinear coefficient, and $z$ is the beam propagation coordinate. Since in our case we estimate that beam propagation involves a significant number of guided modes, we directly numerically solved Eq.\ref{eq:1}, which is computationally much more efficient than using a modal expansion. We used a standard split-step Fourier method with periodic boundary conditions in time $t$; we set the transverse field to zero at the boundaries of the spatial window. We used an integration step of 0.02 mm and a 64x64 grid for a spatial window of $150$ $\mu m \times 150$ $\mu m$. 

In the numerics, we considered a GRIN MMF using the previously mentioned parameter values of $\rho$, $n_{co}$, $n_{cl}$ and $\kappa^ {''}$, and the nonlinear refractive index value $n_2=3.2\times 10^{-20}m^2/W$. 
We took a truncated parabolic profile of the refractive index in the transverse domain $n(x,y)^2=n_{co}^2(1-2\Delta r^2/\rho^2)$ for $r<\rho$ and $n(x,y)=n_{cl}$ otherwise, where $r^2=x^2+y^2$. We did not include the Raman scattering effect, which is justified by the experimental observations that the GPI sidebands appear below the Raman threshold. Moreover, in the simulations we neglected material absorption, because of the relatively short lengths of MMF involved in our experiments. In the numerics we used an input beam diameter of $40$ $\mu m$. To reduce the computational time we used  a  pulse duration of $9$ $ps$, an input intensity of $I=10$ $GW/cm^2$ (i.e., a peak power $P_{p-p}=160$ $kW$) and a fiber's length of 0.4 m.

\begin{figure}
\includegraphics[width=\linewidth]{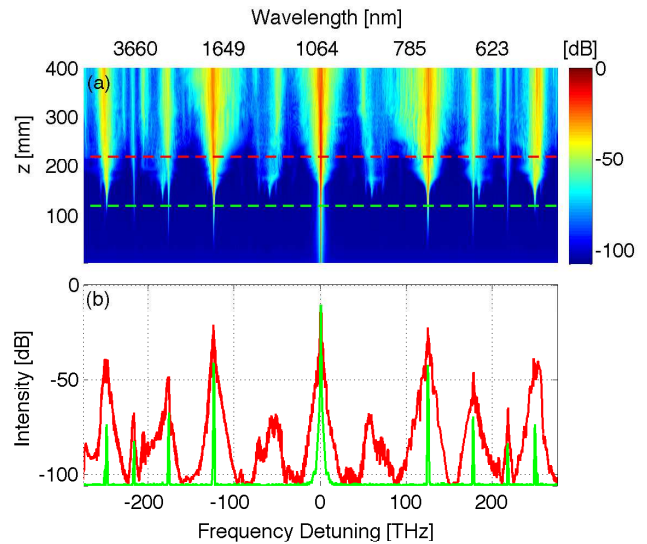}
\caption{Numerical results of (a) spectral evolution upon propagation length $z$ in the GRIN MMF for the pump intensity $I=10$ $GW/cm^2$, and (b) corresponding spectra generated at $z=0.12$ $m$ (green curve) and $z=0.22$ $m$ (red curve) as marked by red and green dashed lines in panel (a), respectively; Intensity is normalized to the peak value along $z$.  \label{fig:2}}
\end{figure}

Figure \ref{fig:1} shows an example of experimental spectrum observed at the input peak power $P_{p-p}$ = 50 kW. As can be seen from Fig.\ref{fig:1}, a series of non-uniformly spaced spectral peaks is present, that covers a remarkably wide visible spectral range starting from 730 nm and then gradually shifting down to 450 nm. 
The black curve in Fig.\ref{fig:1} reports a near-IR portion of the spectrum measured by using another spectrum analyzer able to detect the spectral range from 1200 nm till 2400 nm. The black curve clearly exhibits the presence of the Stokes sideband of the first GPI order, whose frequency shift from the pump is consistent with the corresponding anti-Stokes shift.
Experimental observations of Fig.\ref{fig:1} are quantitatively well reproduced by the numerical simulations illustrated in Fig.\ref{fig:2}. Panel (a) of Fig.\ref{fig:2} shows the spectral evolution (the colormap is in dB scale) as a function of propagation distance $z$ in the MMF, for the pump intensity $I=10$ $GW/cm^2$. Whereas panel (b) of Fig.\ref{fig:2} displays two examples of corresponding output spectra, numerically computed for $z$=0.12 m (green curve) and $z$=0.22 m (red curve). In particular, we can see that GPI first generates a series of narrow spectral sidebands, which subsequently broaden upon propagation, until a spectrum with a shape very close to the experimental results shown in Fig.\ref{fig:1} is obtained. Note that Fig.\ref{fig:2} shows both the visible and the infrared portions of the spectrum. 

Both the theory of GPI and numerical results predict that the entire series of experimental spectral peaks that are observed on the anti-Stokes side of the pump should have their equivalent counterparts on the Stokes side, with an equal set of frequency detunings from the pump. However in practice the spectral region above 2.5 $\mu m$ is expected to be heavily absorbed by the fiber. 
Nevertheless, our experiments show that it may be possible to generate from a quasi-CW laser source an ultra-broadband supercontinuum radiation, coherently linking the entire optical spectrum ranging from the visible until the far infrared domain, by exploiting a single nonlinear effect, the GPI. Note that the experimental observation of additional Stokes spectral peaks was limited by the performance of our visible-near IR spectrum analyzers whose upper limit in wavelength was 2400 nm.

\begin{figure}
\includegraphics[width=\linewidth]{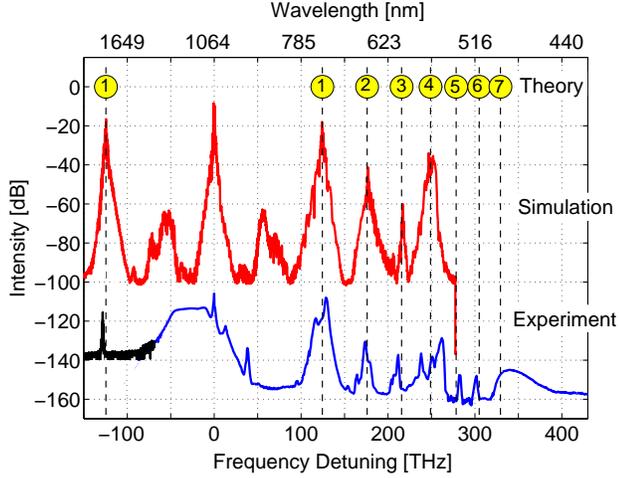}
\caption{Simulated (middle, red curve) and measured (bottom, black-blue curve) series of GPI sidebands; yellow circles (top) show the frequency position of the sidebands calculated analytically. Experimental spectrum (bottom, blue curve) contains higher-order anti-Stokes peaks out of the validity of the theoretical model.  \label{fig:3}}
\end{figure}

In Fig.\ref{fig:3} we compare the experimental GPI sideband frequency positions with their corresponding numerical and analytical values. 
The yellow circles in Fig.\ref{fig:3} represent the analytically predicted frequency positions, spaced by $\sqrt{h}*124.5$ $THz$ with $h=1,2,3,4$. Whereas, the middle red curve and the bottom blue curve show the simulated and the measured spectrum, respectively. As we can see from the figure, there is an excellent agreement between experiments and theory, as far as the spectral position of the GPI sidebands is concerned (until $h=4$). Note that the conversion efficiency of GPI is surprisingly strong, since the intensity of the first GPI anti-Stokes sideband is only 2dB below the residual pump. Additional frequency sidebands can be induced by cascade FWM like those observed around the fourth-order GPI. Whereas, the sideband at 55 THz seen in the numerical spectrum may also be observed in the experiment when increasing the input pump power up to $P_{p-p}=60$ $kW$. 

Next we analyzed the dependence of the GPI sideband frequency positions on the pump power level. Numerical simulations predict that increasing the pump intensity by a factor of 4 (e.g., from $I=5$ $GW/cm^2$ to $I=20$ $GW/cm^2$, or from $P_{p-p}=80$ $kW$ to $P_{p-p}=320$ $kW$) leads to the relatively small (with respect to $f_m=125$ $THz$) nonlinear frequency shift of $\sim$ $2$ $THz$ for the first-order sidebands. The corresponding shift of higher-order sidebands is further reduced by a factor $\sim{h^{-1}}$. Given its small relative value, the pump-power dependence of the sideband shift was hardly observed in our experimental set-up. In fact, the dynamics of our spectrum analyzers and the damage threshold of the fiber allowed us to vary the output power by a factor of 2.5 only, i.e., from $P_{p-p}=30$ $kW$ to $P_{p-p}=74$ $kW$, $i.e.$, below the minimum range of pump power variation that is necessary to induce a detectable GPI sideband frequency shift.

In the experiments, we were also able to observe higher-order anti-Stokes resonance peaks, which are beyond one octave from the pump, that is beyond the range of validity of the NLSE  (equation \ref{eq:1}). Remarkably, as demonstrated by the bottom blue curve in Fig.\ref{fig:3}, the spectral positions of these sidebands fit well with the frequency positions that are calculated by using the analytical formula $\sqrt{h}*124.5$ $THz$, when keeping on with the respective order ($i.e.$ $h=5,6,7$).  

\begin{figure}
\includegraphics[width=\linewidth]{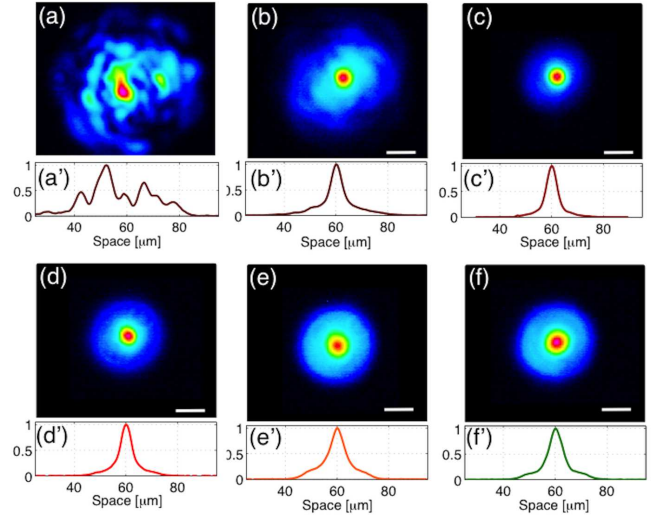}
\caption{Experimental 2D output spatial shapes (intensity in linear scale) and corresponding beam profiles (normalized intensity) versus x (y=0 section) at the pump wavelength (1064 nm) for an input power (a, a') $P_{p-p}=0.06$ $kW$ and  (b, b') $P_{p-p}=50$ $kW$, as well as, at first four orders anti-Stokes sidebands at (c, c') 750 nm, (d, d') 650 nm, (e, e') 600 nm, (f, f') 550 nm, for $P_{p-p}=50$ $kW$. Scale bar: $10$ $ \mu m$ \label{fig:4}}
\end{figure}

Finally we experimentally studied the spatial profile of the output beam shape of the various spectral components. In addition to the pump at 1064 nm, we analyzed the spatial profile of anti-Stokes peaks, by selecting the wavelengths at 750 nm, 650 nm, 600 nm and 550 nm. We used a CCD camera for recording a near-field image of the output face of the fiber with a magnification G=40, and a selection of 10 nm-wide bandpass interference optical filters. Panels (a)-(f) of Fig.\ref{fig:4} show 2D spatial beam distributions at the wavelengths of the pump and the first four orders of the GPI, with their corresponding 1D sections displayed in the panels (a')-(f'). 
At low input powers (for $P_{p-p}=0.06$ $kW$) the output pump exhibits a highly multimode, speckled structure, as shown in panels (a) and (a') of Fig.\ref{fig:4}. However, at the input power $P_{p-p}=50$ $kW$ we strikingly observed that the pump and the GPI sidebands are all carried by 
stable bell-shaped spatial beam profiles, as presented in panels (b)-(f) and (b')-(f') of Fig.\ref{fig:4}.
Note that a multimode low-level pedestal was still sufficient to maintain the self-imaging effect of the GRIN MMF, that is required for GPI sideband generation.

To summarize, in this work we reported the first observation of multiple spectral sidebands, generated by geometric parametric instability in the normal dispersion regime of a multimode optical fiber. In our experiments, the instability is a spatio-temporal effect, whereby intense spectral peaks are induced in the frequency content of a light beam by the longitudinal oscillation of the transverse beam profile. The observed first-order sidebands exhibit a frequency detuning from the pump as large as 123.5 THz. The shape and spectral locations of the individual sidebands are in excellent agreement with theoretical predictions and numerical simulations. Remarkably, we have shown that GPI sidebands are carried by a stable, bell-shaped transverse structure. 

By exploiting a single nonlinear process, our experiment permits to directly frequency convert a near-infrared laser pump into the whole visible spectral range over a bandwidth spanning more than 300 nm. Such type of light source has exciting potential applications to biological imaging. Moreover, our results provide a significant contribution to the emerging field of spatio-temporal multimode nonlinear complex systems, and may have significant consequences in other physical settings where Gross-Pitaevskii equation applies, such as Bose-Einstein condensation.

\begin{acknowledgments}
K.K, A.T., B.M.S., A.B. and V.C. acknowledge the financial support provided by Bpifrance OSEO (Industrial Strategic Innovation Programme), by R\'egion Limousin (C409-SPARC) and ANR Labex SIGMA-LIM. S.W. acknowledges support by the Italian
Ministry of University and Research (MIUR) (2012BFNWZ2). A.B. and G.M. acknowledge support by iXcore research foundation, Photcom R\'egion Bourgogne and ANR Labex Action. The authors thank B. Kibler and P. Tchofo-Dinda for helpful discussions.
\end{acknowledgments}




\bibliography{biblio}

\end{document}